 \documentclass[12pt]{article}                               

\usepackage{amsmath,amssymb,graphics,feynmp}
{\def\$#1: #2 ${#2}}

\begin{document}

\title{%
  Electroweak Event Generators\\
  for LEP2 and the Linear Collider%
 \thanks{
    Invited talk at CRAD'96, 3${}^{\text{rd}}$      
    International Symposium on Radiative            
    Corrections, Cracow, 1996. To appear in the     
    proceedings.}
  }

\author{%
  Thorsten Ohl\thanks{%
       Supported by Bundesministerium f\"ur Bildung,
       Wissenschaft, Forschung und Technologie, Germany.}
    \thanks{e-mail:
      \texttt{Thorsten.Ohl@Physik.TH-Darmstadt.de}} \\
  \hfil \\
  Technische Hochschule Darmstadt \\
  Schlo\ss gartenstr. 9 \\
  D-64289 Darmstadt \\
  Germany}

 \date{
   IKDA 96/21\\                                     
   hep-ph/9610349\\                                 
   October 1996}                                    

\maketitle
\begin{abstract}
  I discuss the state of the art and outline direction for research in
  event generation for electroweak physics at LEP2 and $e^+e^-$~Linear
  Colliders.
\end{abstract}

\begin{fmffile}{crad96pics}
\section{Introduction}
\label{sec:introduction}

With the start of experimentation at LEP1, a new era of precision
measurements in high energy physics had begun.  The potential of such
machines calls for a new level of sophistication in the theoretical
tools.  The ability to measure parameters of the electroweak standard
model at the permille level requires precise theoretical predictions
of observable quantities.

Today's sophisticated multi purpose detectors have a complicated
geometry that can not be described by simple acceptance and efficiency
functions, but require detailed simulation instead.  For this
purpose, Monte Carlo event generators and sophisticated integration
algorithms are indispensable.

The theoretical issues involved in the calculation of differential
cross sections have been covered by other speakers at this
conference~\cite{Beenakker:1996:CRAD96} and need not be repeated
here. I will therefore concentrate on the more technical issues
involved in turning these theoretical predictions into numbers that
can be compared with experimental results.

I will start in section~\ref{sec:mc} with reviewing the motivation
for Monte Carlo methods and I will describe the methods in use today in
section~\ref{sec:12-roads}, including Quasi Monte Carlo in
subsection~\ref{sec:QRNG}.  I will continue with radiative
corrections in section~\ref{sec:RC} and discuss specific problems
with automated calculations of radiative corrections in
subsection~\ref{sec:automatic-RC}.  After reviewing the status of
event generators for LEP2 in section~\ref{sec:LEP2}, I describe the
new challenges at a Linear Collider in section~\ref{sec:LC},
including a discussion of beamstrahlung in
section~\ref{sec:beamstrahlung}.  After a brief look into the future
in section~\ref{sec:futures} and a few bits and pieces in
section~\ref{sec:misc}, I will conclude.

\section{The Need for Monte Carlo}
\label{sec:mc}

The increasing energy of~$e^+e^-$ storage rings and Linear Colliders
opens up channels with many particles in the final state, that are
calculable in the electroweak standard model.  Continuing the
precision tests of the standard model at higher energies requires a
precise calculation of the cross sections for these channels.
In addition, these channels provide backgrounds for new particle
searches, which calls for theoretical control of these cross
sections as well.

Numerical integration estimates the integral of a function~$f$ by a
weighted sum of the functions' values sampled on a set~$\{x\}$ of
points
\begin{equation}
\label{eq:numint}
  \int\!dx\,f(x) = \left\langle f \right\rangle
    = \sum_{x\in\{x\}} w(x) f(x) \,.
\end{equation}
Monte Carlo integration is the special case of~(\ref{eq:numint}), in
which~$\{x\}$ is a sample of pseudo random numbers
and~$w(x)=1/\left|\{x\}\right|$, where~$\left|\{x\}\right|$ denotes
the number of points in~$\{x\}$.  This case is of special importance,
because the law of large numbers guarantees the convergence
of~(\ref{eq:numint}) for any~$f$, only the scale of the error depends
on~$f$.  In addition, the statistical nature of the method allows
reliable error estimates by repeating the evaluation with varying
random sets and checking the Gaussian distribution of the results.

Monte Carlo is the only known method that allows the integration of
differential cross sections for arbitrary final states with arbitrary
phase space cuts.  Monte Carlo event generation is required for
realistic simulations of the acceptance and efficiency of modern
detectors with their complicated geometry.  Thus, Monte Carlo is
\emph{the} universal tool for turning \emph{actions} into measurable
\emph{numbers}.

\subsection{Discovery vs.~Precision Physics}
\label{sec:discovery/precision}

Discovery physics and precision physics call for different approaches
to event generation.  Precision tests of the standard model require
complete calculations, that include radiative corrections and
``background'' diagrams. Since the number of diagrams for many
particle final states is large (up to~144 for four fermions and
thousands for six fermions), such calculations are technically
demanding.  Fortunately, the parameter space is restricted and allows
a comprehensive analysis.

On the other hand, the search for physics beyond the standard model in
general does not need radiative corrections other than the initial
state radiation of photons.  This makes the calculations technically
easier, but the need to cover a lot of models with a vast parameter
space creates other problems.  Since tree level calculations folded
with initial state structure functions are usually sufficient,
computer aided approaches can help to cover the models and their
parameter spaces.

\section{Twelve Roads from Actions to Answers}
\label{sec:12-roads}

There are twelve different methods for getting answers (cross sections
and event rates) from actions (the definition of the physical model in
perturbative calculations).  These can be factorized in three roads
from actions to amplitudes and four ways from amplitudes to answers.

\subsection{The Three Roads from Actions to Amplitudes}
\label{sec:3-roads}

The traditional textbook approach to deriving amplitudes from actions
are \emph{manual calculations}, which can be aided by computer algebra
tools.  The time-honored method of calculating the squared amplitudes
directly using trace techniques is no longer practical for today's
multi particle final states and has generally been replaced by
helicity amplitude methods.  Manual calculation has the disadvantage of
consuming a lot of valuable physicist's time, but can provide insights
which are hidden by the other approaches discussed below.

The currently most successful and increasingly popular technique is to
use a well tested \emph{library of basic helicity amplitudes} for the
building blocks of Feynman diagrams and to construct the complete
amplitude directly in the program in the form of function calls.  A
possible disadvantage is that the differential cross section is nowhere
available as a formula, but the value of such a formula is limited
anyway, since they can hardly be printed on a single page anymore.

\emph{Automatic calculations} are a further step in the same
direction.  The Feynman rules (or equivalent prescriptions) are no
longer applied manually, but encoded algorithmically.  This method
will become more and more important in the future, but more work is
needed for the automated construction of efficient event generators
and for the implementation of radiative corrections (see also
section~\ref{sec:automatic-RC}).

\subsection{The Four Roads from Amplitudes to Answers}
\label{sec:4-roads}

\emph{Analytic and semi-analytic calculations} have the potential to
provide the most accurate results.  Unfortunately, fully analytic
calculations are not feasible for more than three particles in the
final state, that is for most of the interesting physics at LEP2
and the Linear Collider.  Still, semi-analytic calculations, in which
some integrations are performed analytically and the remaining
integrations are done numerically are possible for certain simple cuts
and provide useful benchmarks with an accuracy unmatched by the other
methods.

The most flexible approach is \emph{Monte Carlo integration}, which
converges under very general conditions as
\begin{equation}
\label{eq:mc-convergence}
  \frac{\delta\left\langle f\right\rangle}{\left\langle f\right\rangle}
   \propto
    \frac{1}{\sqrt{N}} \sqrt{\left\langle f^2 \right\rangle
                             - \left\langle f \right\rangle^2}\,,
\end{equation}
with~$N$ denoting the number of function evaluations.
As long as the integration variables can be transformed such that
the integrand~$f$ does not fluctuate too wildly, this method can be
implemented easily and efficiently.  The quadratic increase on
computing resource consumption with the precision puts a practical
limit on the attainable precision, however.

\emph{Event Generation} is a special case of Monte Carlo integration
in which an ensemble of kinematical configurations is generated that
is distributed according to the differential cross section.  Such
ensembles allow the straightforward simulation of the non-perturbative
physics of the fragmentation and hadronization of strongly interacting
particles and of the detector.  If the weight function is bounded,
rejection methods can be used to trivially turn a Monte Carlo code
into an event generator.  Hand tuning is required, however, to make the
code efficient.

The rate of convergence in~(\ref{eq:mc-convergence}) is guaranteed by
the law of large numbers and can only be improved if we turn away from
pseudo random numbers and switch to \emph{deterministic integration}
and \emph{Quasi Monte Carlo}.  Empirical evidence from known quasi
random number sequences suggests that
\begin{equation}
\label{eq:qmc-convergence}
  \frac{\delta\left\langle f\right\rangle}{\left\langle f\right\rangle}
   \propto
    \frac{\ln^n N}{N}
\end{equation}
(cf.~(\ref{eq:D_infty_min}), below)
is possible by using point sets that are more uniform than both random sets
and hyper-cubic lattices in high dimensions.  Quasi Monte Carlo
integration has been applied successfully to four fermion production
at LEP2~\cite{Passarino:1996:WTO}.  Nevertheless, more work is needed,
because there are too few theorems for realistic function spaces.
Phase spaces of varying dimensionality, as in branching algorithms,
have not been studied at all yet.

\subsection{Quasi Random Numbers}
\label{sec:QRNG}

It is intuitively clear that the best convergence will be gotten by
using the most uniform point sets.  It is less obvious how such
uniform point sets look like and how to construct them.  Let us
therefore take a closer look at such point sets.  More detail and
references can be found in~\cite{James/etal:1996:QMC}.

Let us assume for simplicity that we can map the integration region to
the $n$-dimensional unit hypercube: $I = [0,1]^{\times n}$.  Using the
characteristic function
\begin{equation}
  \chi(y|x) = \prod_{\mu=1}^n \Theta(y^\mu - x^\mu)\,,
\end{equation}
we can define the \emph{local discrepancy} of a point set~$\{x\}$ for
each~$y\in I$:
\begin{equation}
  g(y|\{x\}) = \frac{1}{|\{x\}|}
      \sum_{x\in\{x\}} \chi(y|x) - \prod_{\mu=1}^n y^\mu\,.
\end{equation}
Obviously, the local discrepancy measures how uniformly~$\{x\}$ covers
the hypercube.  We can now define various versions of the \emph{global
discrepancy} of the point set~$\{x\}$:
\begin{subequations}
\begin{align}
  D_m(\{x\}) &= \int\!dy\, \left(g(y|\{x\})\right)^m \\
\label{eq:D_infty}
  D_\infty(\{x\}) &= \sup_y |g(y|\{x\})|\,.
\end{align}
\end{subequations}
A lower bound for~$D_\infty$ can be established
\begin{equation}
\label{eq:D_infty_min}
  D_\infty(\{x\})
    \ge C(n) \frac{\ln^{(n-1)/2}|\{x\}|}{|\{x\}|}\,,
\end{equation}
where~$C(n)$ depends only on the dimension of the hypercube and is in
particular independent of~$\{x\}$.
The global discrepancy~$D_2$ of a regular hypercubic lattice
and a random point set can easily be calculated
\begin{subequations}
\begin{align}
  D_2(\{\text{lattice}\})
    &= \frac{n}{4\cdot3^n}\frac{1}{|\{\text{lattice}\}|^{2/n}}
       + \cdots \\
  D_2(\{\text{random}\})
    &= \left(\frac{1}{2^n}-\frac{1}{3^n}\right) \frac{1}{|\{\text{random}\}|}\,.
\end{align}
\end{subequations}
This shows that hypercubic lattices are \emph{less} uniform than random
point sets for more than two dimensions.  This result is not
surprising, however, because we know from experience that Monte Carlo works
better than uniform integration formulae in higher dimensions.

Unfortunately, while it is intuitively obvious that discrepancy and
integration error are related, it is much harder to derive
mathematically rigorous results for realistic integrands, in
particular for those with singularities and discontinuities.  This is
the price to pay for potentially faster convergence and more research is needed.

\section{Radiative Corrections}
\label{sec:RC}

At high energies, electromagnetic radiative corrections are enhanced
by large logarithms~$\ln(s/m^2)$ and precision calculations of the
hard cross section are useless if the radiative corrections are not
under control.

\begin{figure}
  \begin{center}
    \raisebox{30mm}{a)}
    \begin{fmfgraph*}(50,35)
      \fmfpen{thick}
      \fmfleftn{i}{2} \fmfrightn{o}{6}
      \fmfbottom{d1,d2,p,d3}
      \fmflabel{$p$}{i1}
      \fmflabel{$k$}{p}
      \fmf{phantom,tension=3}{i1,v,i2}
      \fmf{fermion}{o1,v,o2}
      \fmf{fermion}{o3,v,o4}
      \fmf{fermion}{o5,v,o6}
      \fmfblob{.2w}{v}
      \fmffreeze
      \fmf{fermion}{i1,vp}
      \fmf{fermion,tension=0.5,label=$1/pk$,label.side=left}{vp,v}
      \fmf{fermion}{v,i2}
      \fmffreeze
      \fmf{photon}{vp,p}
    \end{fmfgraph*}
    \qquad
    \raisebox{30mm}{b)}
    \begin{fmfgraph}(50,35)
      \fmfpen{thick}
      \fmfleftn{i}{2} \fmfrightn{o}{6}
      \fmf{phantom,tension=3}{i1,v,i2}
      \fmf{fermion}{o1,v,o2}
      \fmf{fermion}{o3,v,o4}
      \fmf{fermion}{o5,v,o6}
      \fmfblob{.2w}{v}
      \fmffreeze
      \fmf{fermion}{i1,vp,v,i2}
      \fmffreeze
      \fmf{photon,right}{vp,v}
    \end{fmfgraph}
  \end{center}
  \caption{\label{fig:virtual-real-cancellations}%
    Cancellations of real~(a) and virtual~(b) singularities.}
\end{figure}
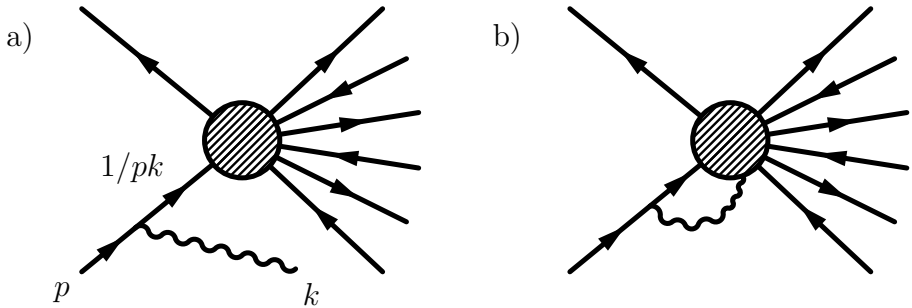

Perturbative calculations at fixed order order in perturbation theory
are not sufficient, because amplitudes for the emission of photons
(figure~\ref{fig:virtual-real-cancellations}a) have
$1/\omega$-soft and $1/\theta$-collinear singularities.  These
singularities are cancelled
\begin{equation}
  \frac{1}{\omega} \to
  \left(\frac{1}{\omega}\right)_+ = \lim_{\epsilon\to0} \left[
  \frac{1}{\omega} \Theta(\omega-\epsilon) \Bigr|_{\text{real}}
    + \ln\epsilon \cdot \delta(\omega) \Bigr|_{\text{virtual}} \right]
\end{equation}
in the cross section by opposite-sign singularities in the virtual
corrections (figure~\ref{fig:virtual-real-cancellations}b)
to the indistinguishable process without emission.  While
sufficiently integrated cross sections remain positive, the
differential cross section without emission must be negative for the
cancellation to take effect.  Only after resummation of the
perturbation series, these singularities become integrable
like~$1/\omega^{1-\beta}$ and result in strictly positive differential
cross sections.  The combination of today's collider energy and of the
energy- and angular resolution of today's detectors have made this
resummation a practical necessity.

The most popular approach today is the summation of the leading
logarithmic initial state contributions through structure functions
\begin{equation}
\label{eq:ISR}
   \frac{d\sigma}{d\Omega} (s) =
      \int\!dx_+dx_-\,
         \left|\frac{\partial\Omega'}{\partial\Omega}(x_+,x_-)\right|
              D(x_+,s) D(x_-,s) \frac{d\sigma^0}{d\Omega'} (s')\,.
\end{equation}
The structure functions~$D$ can be obtained from finite order
perturbation theory with resummed (``exponentiated'') singularities,
from Yennie-Frautschi-Suura summation, from the DGLAP evolution
equations or from parton shower algorithms.

A drawback of this approach is that the transversal momentum carried
away by the photons is ambiguous in leading logarithmic approximation.
For practical purposes, most
authors use the $1/(p\cdot k)$~behaviour suggested by the most
singular pieces.
This problem is worrisome for experiments that measure the amount
of initial state radiation by tagging photons in order to select an
event sample with effective center-of-mass energy below the nominal
machine energy.
Theoretically most satisfactory are algorithms that match the hard
matrix element outside the singular region to the resummed cross
section in the singular region.

Another universal correction appears in the $W^+W^-$-production close
to threshold at LEP2.  The long range Coulomb interaction of
slow~$W$'s gives rise to a substantial change of the cross section.
This correction is easy to implement and available in most computer
codes.

The issue of QCD corrections was the subject of some debates during
the 1995 LEP2 workshop.  The ``naive'' QCD correction
\begin{equation}
  \left( 1 + \frac{\alpha_{QCD}}{\pi} \right)
     \cdot \Gamma(W/Z\to q\bar q')
\end{equation}
is obtained by integrating over the gluon in the decay of a vector
boson into quark pairs.  The effect on the width and the branching
ratios is substantial and must be included in some way in the
calculations.  At the same time, this naive approach is theoretically
questionable, because the unavoidable experimental cuts will
invalidate the fully inclusive calculation.  Since there are again
cancellations between real and virtual contributions at work, this is
not a priori a numerically small effect.  Furthermore, the resonant
diagrams do not form a gauge invariant subset and a more complete
analysis is called for.

Such studies are under way~\cite{Maina/etal:1996:CC10_O(alpha)}.  But
until their results are included in the standard computer codes,
the naive QCD corrections are available in most programs because the
effect is substantial.

Weak radiative corrections are certainly more interesting from the
physics perspective, but they will hardly be measurable at LEP2.  
Their calculation is very tough, because effective Born approximations
similar to those successful at LEP1, do \emph{not} exist for
differential cross sections in $W^+W^-$~pair production.  Since
complete calculations  do not yet exist except for the on-shell
production of stable~$W$'s, gauge dependent \emph{ad-hoc} resummations
of the propagators are needed.

For practical purposes at LEP2, the gauge dependence problem has been
solved by the fermion loop scheme~\cite{Argyres/etal:1995:BHF}, but
more complete investigations and implementations remain desirable,
because stronger effects are expected at the Linear Collider.

\subsection{Radiative Corrections in Automatic Calculations}
\label{sec:automatic-RC}

Radiative corrections in gauge theories pose a particular problem for
automatic calculations~(see \cite{Kato:1996:CRAD96} for a general
review of automatic calculations).
Formally~\cite{Caravaglios:1996:CRAD96}, loops can be incorporated in
existing tree level generators by replacing the classical
action~$S(A,\psi,\bar\psi)$ with the generating
functional~$\Gamma(A,\psi,\bar\psi)$ of one particle irreducible
diagrams.  Unfortunately, this approach works only for loops
consisting of heavy particles.

As mentioned above, the amplitude for the emission of massless
particles~(figure~\ref{fig:virtual-real-cancellations}a) contains
infrared and collinear singularities, which are cancelled in the cross
section by the virtual
contributions~(figure~\ref{fig:virtual-real-cancellations}b) for
degenerate final states.  The problem is that both contributions have
to be regularized and that the appropriate regularization of the loop
in diagram~\ref{fig:virtual-real-cancellations}b depends on which
external leg it is attached to.  Therefore it is not possible to
describe the loop diagrams by a position independent generating
functional~$\Gamma(A,\psi,\bar\psi)$.  More sophisticated algorithms
which analyze to topology of the graphs are needed.  A new approach to
this problem for QCD jet cross sections has been presented at this
conference~\cite{Catani:1996:CRAD96}.

Until this problem is solved, automatic calculations are useful for
the emission of hard photons which are separated from all charged
photons.  Integrated cross sections can be calculated in the
leading logarithmic approximation only by folding hard cross sections
with structure functions~(\ref{eq:ISR}).

\section{Status of Event Generators for LEP2}
\label{sec:LEP2}

\begin{table}
  \let\gbox\relax
  \small
  \begin{center}
  \renewcommand{\tabcolsep}{3pt}
\begin{tabular}{|c||c|c|c|c|c|c|c|c|c|}\hline
  Program           &Type&Diag.  &ISR    &FSR&NQCD&Clb.&AC &$m_f$&Jets\\\hline
  \hline
  \texttt{ALPHA}    &MC  &all    &$-$    &$-$&$-$ &$-$ &$-$&$+$  &$-$ \\\hline
  \texttt{COMPHEP}  &EG  &all    &SF     &$-$&$ $ &$-$ &$-$&$+$  &$-$ \\\hline
  \texttt{ERATO}    &MC  &all    &SF     &$-$&$+$ &$-$ &$+$&$-$  &$+$ \\\hline
  \texttt{EXCALIBUR}&MC  &all    &SF     &$-$&$+$ &$+$ &$+$&$-$  &$-$ \\\hline
  \texttt{GENTLE}   &SA  &NCC    &SF/FF  &$-$&$+$ &$+$ &\gbox{$+$}
                                                       &PS   &$-$ \\\hline
  \texttt{GRC4F}    &EG  &all    &SF/PS  &PS &$+$ &\gbox{$+$}
                                                       &\gbox{$+$}
                                                           &$+$  &$+$ \\\hline
  \texttt{HIGGSPV}  &EG  &NC$nn$ &SF     &$-$&    &n/a &$-$&$\pm$&$-$ \\\hline
  \texttt{KORALW}   &EG  &all    &YFS    &SF &$+$ &$+$ &$-$&$\pm$&$+$ \\\hline
  \texttt{LEPWW} ($\dagger$)
                    &EG  &CC03   &$\mathcal{O}(\alpha)$
                                 &$+$&$+$&$-$ &$+$&$-$  &$+$ \\\hline
  \texttt{LPWW02}   &EG  &CC03   &SF     &SF &$+$ &$+$ &$-$&$\pm$&$+$ \\\hline
  \texttt{PYTHIA}   &EG  &CC03   &SF$+$PS&PS &$+$ &$+$ &$-$&$\pm$&$+$ \\\hline
  \texttt{WOPPER}   &EG  &\gbox{CC11} 
                                 &PS     &$-$&$+$ &$+$ &\gbox{$+$}
                                                           &$\pm$&$+$ \\\hline
  \texttt{WPHACT}   &\gbox{EG} 
                         &all    &SF     &$-$&$+$ &$+$ &\gbox{$+$}
                                                           &$+$  &\gbox{$+$}
                                                                      \\\hline
  \texttt{WTO}      &Int.&NCC    &SF     &$-$&$+$ &$+$ &$-$&$-$  &$-$ \\\hline
  \texttt{WWF}      &EG  &\gbox{CC20} 
                                 &SF$+$ME&ME &$+$ &$+$ &$+$&$+$  &$+$ \\\hline
  \texttt{WWGENPV}  &EG  &\gbox{CC20}
                                 &\gbox{SF${}_{p_T}$}
                                         &\gbox{SF${}_{p_T}$}
                                             &$+$ &$+$ &$-$&$\pm$&$+$ \\\hline
\end{tabular}
  \end{center}
  \caption{\label{tab:LEP2MC}%
    Properties of the available computer codes for $W^+W^-$-physics at
    LEP2.  In the `Type' column `MC' stands for Monte Carlo
    integration, `EG' for event generation, `SA' for semi-analytic and
    `Int.' for deterministic integration.  Implemented subsets of
    diagrams are denoted by `CC03' for doubly resonant $W^+W^-$,
    `CC11' for singly resonant $W^+W^-$, `CC20' for final states
    including electrons or positrons, `NC$nn$' for various neutral
    current diagrams and `NCC' for various neutral and charged current
    diagrams. The implementation of initial state radiation is
    denoted by `SF' for structure functions, `FF' for flux functions,
    `PS' for parton showers, `YFS' for Yennie-Frautschi-Suura,
    `ME' for matrix element and `$\mathcal{O}(\alpha)$' for one photon
    bremsstrahlung.  The `NQCD' column applies to naive inclusive QCD
    corrections.  For fermion masses, `$\pm$' denotes
    massless matrix elements with massive kinematics.
    See~\cite{Bardin/etal:1996:WWEG} for references and more details.}
\end{table}

The event generators available for $WW$-physics at LEP2 have been
described in~\cite{Bardin/etal:1996:WWEG}.  An updated summary of the
properties of the available codes is presented in
table~\ref{tab:LEP2MC}.

In~\cite{Bardin/etal:1996:WWEG}, the predictions for LEP2 have been
compared in detail.  In a ``tuned'' comparison with a prescribed
calculational scheme, the implementations have been tested and the
modern dedicated~$e^+e^-\to4f$ codes agreed at a level far better than
required at LEP2.  In a second ``unleashed'' comparison, the different
theoretical approaches of the codes have been compared and the results have
shown that the predictions are under control at the level required for
LEP2.

\section{Towards the Linear Collider}
\label{sec:LC}

We have to wait another decade until a Linear Collider will be
available for experimentation. Nevertheless, the design of the
interaction region and of detectors needs firm predictions for the
expected physics~\cite{Murayama/Peskin:1996:LC_review}
\emph{today}.

Regarding two gauge boson physics (which for precision measurements is
really four fermion physics), most of the LEP2 Monte Carlos in
table~\ref{tab:LEP2MC} can simply be run at higher energies, provided
that they can be interfaced to beamstrahlung codes, as discussed in
subsection~\ref{sec:beamstrahlung} below.

It is unreasonable to expect deviations of the three gauge boson
couplings from the Standard Model values \emph{(anomalous couplings)}
that will be measurable at LEP2 (though this assertion has to be
checked anyway).  This will change at the Linear Collider, because
reasonable values of~$\mathcal{O}(1/(16\pi^2))$ become accessible and
event generators have to support anomalous couplings.  Fortunately,
the majority of the event generators supports them today and
preliminary cross checks have been satisfactory.

As mentioned above, weak radiative corrections will also become
relevant.  Here more work is still needed.

\subsection{$e^+e^-\to 6f + n\gamma$}

At a 500~GeV Linear Collider interesting $e^+e^-\to 6f$~channels open
up.  For the first time, precision measurements of
$t\bar t$~production at a $e^+e^-$-collider will be possible.  An event
generator at the signal diagram level, including bound state effects,
is available.  The study of background diagrams will be necessary and a
cross check of the generator will be desirable.

\begin{figure}
  \begin{center}
    \begin{fmfgraph*}(50,35)
      \fmfpen{thick}
      \fmfleft{ep,em} \fmfright{nub,f1,f2,f3,f4,nu}
      \fmflabel{$e^-$}{em} \fmflabel{$e^+$}{ep}
      \fmflabel{$e^+/\bar\nu$}{nub} \fmflabel{$e^-/\nu$}{nu}
      \fmflabel{$\bar f_1$}{f1} \fmflabel{$f_2$}{f2}
      \fmflabel{$\bar f_3$}{f3} \fmflabel{$f_4$}{f4}
      \fmf{fermion,t=3}{em,wm} \fmf{fermion}{wm,nu}
      \fmf{fermion}{nub,wp} \fmf{fermion,t=3}{wp,ep}
      \fmf{boson}{wp,v,wpd} \fmf{boson}{wm,v,wmd}
      \fmf{fermion}{f1,wmd,f2} \fmf{fermion}{f3,wpd,f4}
      \fmfdot{wm,wp,wmd,wpd} \fmfblob{(.1w)}{v}
    \end{fmfgraph*}
  \end{center}
  \caption{\label{fig:VV->VV}%
    $VV\to VV$ scattering.}
\end{figure}
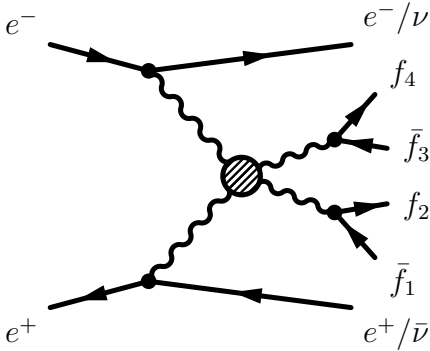
At high energies, vector boson
scattering~(cf.~figure~\ref{fig:VV->VV}), becomes an interesting
$e^+e^-\to 6f$~channel as a probe of the electroweak symmetry breaking
sector.  Work in this direction has started this year.

The general case of~$e^+e^-\to 6f$ is a formidable computational
challenge, because a huge number of diagrams has to be calculated.
At the moment four approaches are being discussed:
\begin{enumerate}
  \item start from on-shell~$e^+e^-\to VVV$, $VV\to VV$ and
    $e^+e^-\to t\bar t$~Monte Carlos and add $V\to f\bar f'$~decays in
    a second step.  This approach is probably not useful for obtaining
    a complete calculation in the end, but it can provide reasonable
    descriptions of the most important, resonant contributions soon.
  \item extent the \texttt{EXCALIBUR}~\cite{Bardin/etal:1996:WWEG}
    algorithm for massless
    four fermion production to six fermions.  It is still unclear how
    to deal with quark masses efficiently in this approach.
  \item start with complete calculations of specific final states for
    which the number of diagrams is manageable.  Work in this direction
    has started and is showing first promising
    results~\cite{Ballestrero:1996:private}.
  \item perform a fully computerized calculation.  Work in this
    direction using the \texttt{GRACE}~\cite{Bardin/etal:1996:WWEG}
    system has started as well. 
\end{enumerate}
These projects will certainly keep the aficionados of standard model
calculations entertained for some years.

\subsection{Beamstrahlung}
\label{sec:beamstrahlung}

The experimental environment at the Linear Collider causes a new
phenomenon that event generators have to deal with.  The largest
(two fermion) standard model cross sections are of the order
\begin{equation}
  \frac{4\pi}{3}\frac{\alpha^2}{s}
    \approx \frac{100\text{fb}}{(\sqrt{s}/\text{TeV})^2}\,,
\end{equation}
and four and six fermion cross sections are suppressed by further factors
of~$\alpha$.  It is therefore clear that interesting physics at the
Linear Collider will require large luminosities of the order of
\begin{equation}
  10^{34}\text{cm}^{-2}\text{sec}^{-1}
     \approx 100 \text{fb}^{-1}\upsilon^{-1}\,,
\end{equation}
where $\upsilon=10^7\text{sec} \approx \text{year}/\pi$ corresponds to
one ``effective'' year of running.  These luminosities can only be
achieved with extremely dense bunches of particles.

Such bunches experience a strong electromagnetic interaction with
non-trivial consequences.  A desired effect of this interaction for
oppositely charged particles is the ``pinch effect'', which increases
the luminosity by further collimating the bunches through the
reciprocal attraction. But the same physics gives rise to undesirable
side effects as well.  The deflection of the particles in the bunches
causes them to loose several percent of their energy as
synchrotron radiation \emph{(beamstrahlung)}.  Therefore, the 
colliding particles will have a non-trivial energy spectrum.  This
spectrum has to be included in the event generators for realistic
predictions. At the same time, the radiated photons take part in
$\gamma e^\pm$- and $\gamma\gamma$-collisions and therefore their
energy spectrum has to be known as well.

Quantitatively, the effect of the beamstrahlung is of the same order
as the effect of initial state radiation.  But unlike ordinary
bremsstrahlung, beamstrahlung can not be treated by ordinary
perturbation theory, because the underlying physics is the interaction
of a particle with \emph{all} the particles in the colliding bunch.
Approximate analytical treatments of this collective effect exist, but
full simulations~\cite{Chen/etal:1995:CAIN} show
that they are not adequate.

\begin{table}
  \begin{center}
    \renewcommand{\arraystretch}{1.1}
    \begin{tabular}{|c||c|c|c|}\hline
          & \texttt{SBAND} & \texttt{TESLA} & \texttt{XBAND}
      \\\hline\hline
      $E/\text{GeV}$                  & 250    &  250    & 250    \\\hline
      $N_{\text{particles}}/10^{10}$  &   1.1  &    3.63 &   0.65 \\\hline
      $\epsilon_x/10^{-6}\text{mrad}$ &   5    &   14    &   5    \\\hline
      $\epsilon_y/10^{-6}\text{mrad}$ &   0.25 &    0.25 &   0.08 \\\hline
      $\beta^*_x/\text{mm}$           &  10.98 &   24.95 &   8.00 \\\hline
      $\beta^*_y/\text{mm}$           &   0.45 &    0.70 &   0.13 \\\hline
      $\sigma_x/\text{nm}$            & 335    &  845    & 286    \\\hline
      $\sigma_y/\text{nm}$            &  15.1  &   18.9  &   4.52 \\\hline
      $\sigma_z/\mu\text{m}$          & 300    &  700    & 100    \\\hline
      $f_{\text{rep}}$                &  50    &    5    & 180    \\\hline
      $n_{\text{bunch}}$              & 333    & 1135    &  90    \\\hline
    \end{tabular}
  \end{center}
  \caption{\label{tab:machines}%
    Three prototypical linear collider designs at~500~GeV:
    \texttt{SBAND} and \texttt{TESLA} are DESY's room temperature and
    superconducting options, \texttt{XBAND} is for KEK's and SLAC's
    plans.}
\end{table}

The full simulations consume too much computer time and memory for
directly interfacing them to particle physics Monte Carlos.  Also, the
input parameters collected in table~\ref{tab:machines} not familiar to
most particle physicists.  The pragmatical solution of this problem is
to describe the result of the simulations by a simple \emph{ansatz},
that captures the essential features.

\begin{figure}
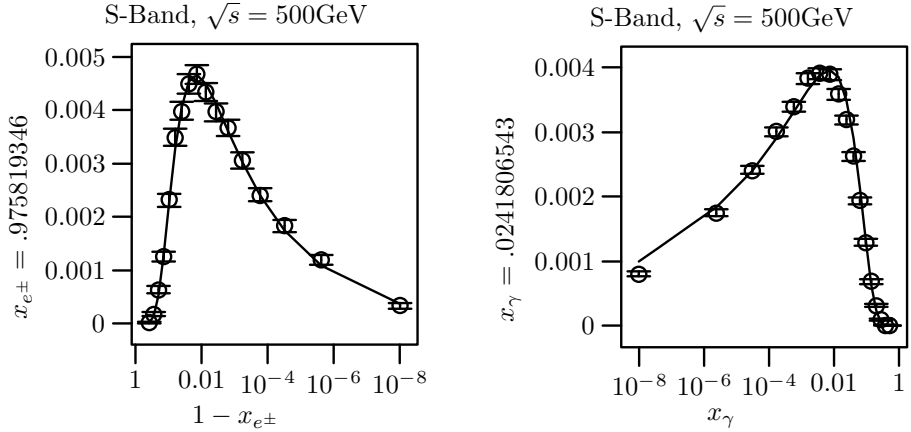

\begin{center}
  \includegraphics{fit.11}\qquad
  \includegraphics{fit.21}
\end{center}
  \caption{\label{fig:circe-fit}%
    Quality of the fits for the TESLA design at~500~GeV and~1~TeV.}
\end{figure}

The approximate solutions motivate a factorized \emph{ansatz}, which
should positive and have \emph{integrable} singularities
at the endpoints~$x_{e^\pm}\to1$ and~$x_\gamma\to0$
\begin{align}
  D_{p_1p_2} (x_1,x_2,s) &= d_{p_1} (x_1)  d_{p_2} (x_2)\\
  d_{e^\pm} (x) &= a_0 \delta(1-x) + a_1 x^{a_2} (1-x)^{a_3} \\
  d_\gamma (x) &= a_4 x^{a_5} (1-x)^{a_6}\,.
\end{align}
Figure~\ref{fig:circe-fit} shows that this ansatz works surprisingly
well.  It has therefore been made available as distribution
functions and non-uniform random number generators in the
\texttt{circe} library~\cite{Ohl:1996:circe}.

\begin{figure}
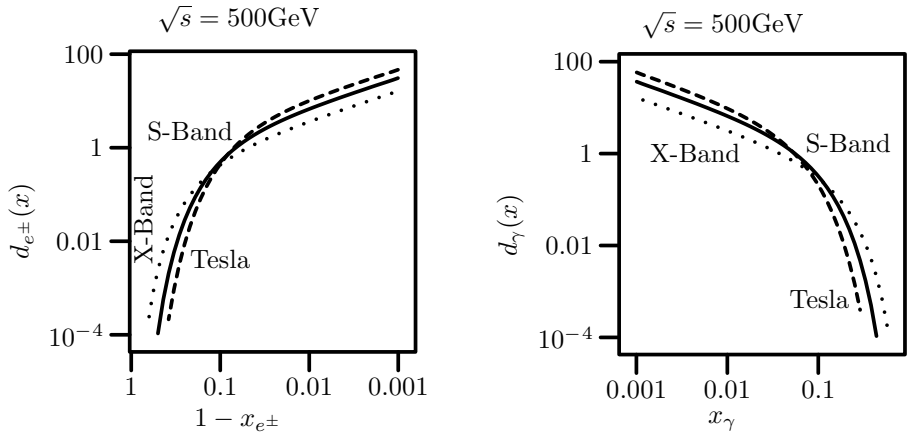

  \begin{center}
    \includegraphics{dist.1}\qquad
    \includegraphics{dist.2}
  \end{center}
  \caption{\label{fig:circe-dist}%
    Results of simulations and fits for the three prototypical linear
    collider designs at~500~GeV.}
\end{figure}

The resulting distributions are displayed in
figure~\ref{fig:circe-dist}, which highlights the substantial differences
in the designs.

\section{Monte Carlo Futures}
\label{sec:futures}

The craft of Monte Carlo event generation for precision physics at
high energy $e^+e^-$-colliders is practised successfully by many
physicists today, as witnessed by the many high quality computer codes
available for LEP2.  Still there are areas where technical progress is
desirable.  

The most promising direction are \emph{generator generators}, where
amplitudes and event generation or integration algorithms are
constructed algorithmically by a computer program from a Lagrangian
for each desired final state.  The systems available today have come a
long way~\cite{Kato:1996:CRAD96}, but are still far from perfect.

While the combinatorics of the generation of the amplitude is
implemented in several systems, most of these implementations do not
scale well to final states with more than four particles.  The
number of diagrams will scale like~$n!$ with the number~$n$ of
particles.  For small~$n$, the resource
consumption of this factorial growth can be matched by faster hardware
in the future, but this is not obvious for the numerical stability of
the code.
Gauge theories have a good high energy behaviour because of strong
cancellations among individual diagrams.  The numerical problems
resulting from these cancellations have to be controlled.  This is
only possible with more sophisticated algorithms that regroup and
partially factorize the amplitudes.  Such algorithms are non trivial
and will need further research.  Here progress is particularly
important for radiative corrections.

A second area of research are adaptive methods for event generation
and integration.  Except for special cases, where the singularities in
the amplitudes are known in advance, today's systems still need human
intervention for finding optimal phase space variables that minimize
the sampling error.  Here more sophisticated algorithms are needed as
well.

Technical advances in this direction will allow to efficiently produce
reliable Monte Carlo codes for precision physics
at high energy Linear Colliders.  Hopefully, this will also propel the
state of the art of event generation for discovery physics to the same
level.

Another promising direction for research is the investigation of Quasi
Monte Carlo methods for event generation and integration.  Here more
experience with realistic applications is needed.

There are however some areas where progress will be slow.  In particular
the interface of the perturbative precision calculations with the non
perturbative simulation of strongly interacting final states is poorly
understood.  A lot of the sophistication in the calculation of
interfering contributions is lost when the perturbative amplitude is
matched to the classical simulation of the fragmentation and
hadronization process.  At LEP2, the problems of color reconnection
and Bose-Einstein correlations are the limiting factor for the
$W$-mass measurements.   When LEP2 data become plentiful, they might
help to improve phenomenological models and give a better control of
the systematical error.

\section{Bits and Pieces}
\label{sec:misc}

Before concluding, I want to take this opportunity to advertize a
welcome addition to the family of pseudo random number generators
available for Monte Carlo integration and event generation.

It is well known that bad random number generators will spoil any
simulation, while (some) good random number generators can consume
macroscopic fractions of the total computer time.

Recently, Donald Knuth made errata~\cite{Knuth:1996:AOCP2_errata} for
his textbook available, which contain a gem of a portable generator.
It is an extremely fast implementation of a lagged Fibonacci
generator~$X_j = (X_{j-100} - X_{j-37}) \mod 2^{30}$, which is
portable even for systems which, like \texttt{FORTRAN}, offer no
unsigned 32-bit arithmetic.  The generator passes all statistical tests, even
(with a slight speed penalty) the birthday spacings test.  But the
most interesting property is the innovative initialization algorithm
for which one can \emph{prove} that it will generate $2^{30}-2$
\emph{statistically independent} sequences from a simple $30$-bit
integer seed.  For most other generators, the statistical independence
for different seeds is much harder to control.

The practical consequence is that Knuth's new algorithm allows
parallel execution and reliable Monte Carlo estimation of errors,
without having to worry about the statistical independence of the
generated samples.
 
\section{Conclusions}
\label{sec:conclusions}

The Monte Carlo codes for precision physics at LEP2 physics are in
excellent shape and the craft of their construction is well
understood.  Tools for the design of the Linear Collider are 
already available, but there is work left to do to make them as
comprehensive as the tools available for LEP2.

Technical progress is still needed for streamlining the computational
approaches that will eventually allow us to investigate the nature of
electroweak symmetry breaking at the Linear Collider.


\end{fmffile}
\end{document}